\documentclass[epj]{svjour}
\usepackage{graphics}
\usepackage{multicol}
\usepackage{graphicx}
\usepackage{booktabs}
\usepackage{amssymb,bm,mathrsfs,bbm,amscd}
\usepackage[tbtags]{amsmath}
\usepackage{lastpage}
\usepackage{changes}
\usepackage{dcolumn}
\usepackage{longtable}
\usepackage[numbers,sort&compress]{natbib}
\usepackage{graphicx}
\usepackage{dcolumn}
\usepackage{bm}
\usepackage{amsmath}
\usepackage{flushend}
\usepackage{changes}
\usepackage[colorlinks,linkcolor=blue,anchorcolor=blue,citecolor=blue,urlcolor=blue]{hyperref}
\usepackage{threeparttable}

\usepackage{array}
\newcommand{\PreserveBackslash}[1]{\let\temp=\\#1\let\\=\temp}
\newcolumntype{C}[1]{>{\PreserveBackslash\centering}p{#1}}
\newcolumntype{R}[1]{>{\PreserveBackslash\raggedleft}p{#1}}
\newcolumntype{L}[1]{>{\PreserveBackslash\raggedright}p{#1}}

\begin{document}
\title{New Geiger--Nuttall law for proton radioactivity}
\author{Jiu-Long Chen\inst{1}\and Jun-Yao Xu\inst{1}\and Jun-Gang Deng\inst{1}\and Xiao-Hua Li\inst{1,2,3,}\thanks{\emph{e-mail:} lixiaohuaphysics@126.com }\and Biao He\inst{4}\and Peng-Cheng Chu\inst{5,}\thanks{\emph{e-mail:} kyois@126.com }%
}                     
%
%
\institute{School of Nuclear Science and Technology, University of South China, Hengyang 421001, China \and Cooperative Innovation Center for Nuclear Fuel Cycle Technology $\&$ Equipment, University of South China, Hengyang 421001, China \and Key Laboratory of Low Dimensional Quantum Structures and Quantum Control, Hunan Normal University, Changsha 410081, China \and College of Physics and Electronics, Central South University, Changsha 410083, China \and School of Science, Qingdao Technological University, Qingdao 266000, China}
\date{Received: date / Revised version: date}
%
\abstract{
In the present work considering the contributions of the daughter nuclear charge and the orbital angular momentum taken away by the emitted proton, we propose a two--parameter formula of new Geiger--Nuttall law for proton radioactivity. A set of universal parameters of this law is obtained by fitting 44 experimental data of proton emitters in the ground state and isomeric state. The calculated results can reproduce the experimental data well. For a comparison, the calculations performed using other theoretical methods, such as UDLP proposed by Qi ${et\ al.}$ [\href{https://journals.aps.org/prc/abstract/10.1103/PhysRevC.85.011303}{Phys. Rev. C \textbf{85}, 011303(R) (2012)}], the CPPM--Guo2013 analyzed by our previous work [Deng ${et\ al.}$, \href{https://link.springer.com/article/10.1140/epja/i2019-12728-0}{Eur. Phys. J. A \textbf{55}, 58 (2019)}] and the modified Gamow--like model proposed by us [Chen ${et\ al.}$, \href{https://iopscience.iop.org/article/10.1088/1361-6471/ab1a56}{J. Phys. G: Nucl. Part. Phys. \textbf{96}, 065107 (2019)}] are also included. Meanwhile, we extend this new Geiger--Nuttall law to predict the proton radioactivity half-lives for $51 \leq Z \leq 91$ nuclei, whose proton radioactivity is energetically allowed or observed but not yet quantified in NUBASE2016.
} 

\maketitle

\section{Introduction}
\label{section 1}

The study of exotic nuclei far away from the $\beta$--stability line has become a very interesting topic both from the experimental and theoretical points of view  with continuous development of the radioactive ion beam facilities. The investigation of exotic nuclei has led to the discovery of a new form of radioactivity---proton radioactivity. In 1960, Goldansky \cite{GOLDANSKY1960482} reported the limits of stability of neutron-deficient isotopes with respect to proton radioactivity. The proton radioactivity was firstly observed in an  isomeric state of $^{53}$Co in 1970 by Jackson et al. \cite{JACKSON1970281,CERNY1970284}. Subsequently, Hofmann et al. and Klepper et al. detected the proton emission from nuclear ground states in $^{151}$Lu \cite{Hofmann1982} and $^{147}$Tm \cite{Klepper1982} independently. Up to now, there are about 29 proton emitters decaying from their ground states and 15 different nuclei choosing to emit protons from their isomeric states, which have been identified between Z = 53 and Z = 83 \cite{SONZOGNI20021,PhysRevLett.96.072501,PhysRevC.72.051601,BLANK2008403,Zhang_2010,Qian2016,PhysRevC.96.034619,Budaca2017}. A detailed account on the summary of the experimental information on ground and isomeric state proton emitters is reported by Sonzogni in Ref. \cite{SONZOGNI20021} and by Blank and Borge in Ref. \cite{BLANK2008403}.

The proton radioactivity can be processed by the Wentzel-Kramers-Brillouin (WKB) approximation method since this process can be treated as a simple quantum tunneling effect through a potential barrier. Based on this method, a lot of models have been proposed to deal with the proton radioactivity such as Woods--Saxon--type potential \cite{PhysRevC.23.920,PhysRevC.45.1688,ALAVI201849}, the effective interactions of density--dependent M3Y (DDM3Y) \cite{Basu_2004,BHATTACHARYA2007263,Yi_Bin_2010}, Jeukenne, Lejeune and Mahaux (JLM) \cite{BHATTACHARYA2007263}, the unified fission model (UFM) \cite{PhysRevC.71.014603,Jian_Min_2010}, the generalized liquid--drop model (GLDM) \cite{PhysRevC.79.054330,Zhang_2010}, the single--folding model (SFM) \cite{PhysRevC.72.051601}, the modified two--potential approach (MTPA) \cite{Qian2016,jlchen257}, the Gamow--like model (GLM) \cite{Zdeb2016,Chen_2019} and the Coulomb and proximity potential model (CPPM) \cite{GUO201354,PhysRevC.96.034619,Deng2019}. For more details about different theories of proton radioactivity, the readers are referenced to Ref. \cite{DELION2006113}. Since the half--life of the proton radioactivity strongly depends on the decay energy $Q_p$ as well as the orbital angular momentum $l$ taken away by the emitted proton, it is indispensable for the process of proton radioactivity to investigate the effects of angular momentum such as being a centrifugal energy term in models or a pivotal correlation in empirical formulas. Based on this theoretical basis, several of empirical formulas included different forms of $Q_p$ and $l$ are also proposed to calculate the half--life of proton radioactivity \cite{PhysRevLett.96.072501,Budaca2017,PhysRevC.85.011303,Zhang_2018,Sreeja2018}. In this work, based on the Geiger-Nuttal(G--N) law \cite{doi:10.1080/14786441008637156} we propose an analytic formula that is more convenient for using in the investigation of proton radioactivity. This formula is the relation of proton radioactivity half--life $T_{1/2}$, decay energy $Q_p$, the charge of the daughter nucleus $Z_d$ and orbital angular momentum $l$. While the parameters are obtained by fitting the experimental data of proton radioactivity.

This article is organized as follows. The details of empirical linear equations, spin--parity selection rule and Q value are  given in Section \ref{section 2}. The new G--N law, detailed calculations and discussion are presented in Section \ref{section 3}. Finally, a brief summary is given in Section \ref{section 4}.

\section{The theoretical method}
\label{section 2}
\subsection{Empirical linear equations}
The first significant correlation between the half--life of $\alpha$ decay process and the decay energy $Q_\alpha$ of the emitted $\alpha$ particle was found by Geiger and Nuttall \cite{doi:10.1080/14786441008637156}, and the Geiger--Nuttall law is written as 
\begin{equation}\label{GNLAW}
\text{log}_{10}T_{1/2}(s)=aQ_{\alpha}^{-1/2}+b.
\end{equation}
Where $a$ and $b$ are the two isotopic chain--dependent parameters of this formula. In order to extend the G--N law to non-isotopes, Brown \cite{PhysRevC.46.811} proposed a universal scaling law with determined the contribution of the charge of the daughter nucleus for $\alpha$ decay half lives of even--even parents $(l = 0$ decays$)$. It is expressed as
\begin{equation}
\text{log}_{10}T_{1/2}(s)=aZ_{d}^{0.6}Q_{\alpha}^{-1/2}+b,
\end{equation}
where $Z_d$ is the charge number of the daughter nucleus.
However, since most proton emitters have different proton numbers and the emitted proton takes nonzero orbital angular momentum, the above two empirical formula can not be fully applicable to the study of proton radioactivity.

In 2006, Delion ${et\ al.}$ \cite{PhysRevLett.96.072501} proposed the first empirical formula for the reduced half--life of proton radioactivity in proton decay processes. The slope parameter is the Coulomb parameter with the charge of the daughter nucleus. This formula is divided into two ranges and written as
\begin{equation}\label{first}
\text{log}_{10}T^k_{\rm{red}}(s)=a_k(\chi-20)+b_k,
\end{equation}
here the parameters are as follows
\begin{equation}
\begin{split}
&a_1=1.31,\ \ b_1=-2.44,\ \  \text{for}\ Z_d<67,\\
&a_2=1.25,\ \ b_2=-4.71,\ \  \text{for}\ Z_d>67.
\end{split}
\end{equation}
Where $T_{\rm{red}}$ is the reduced half--life corrected by the centrifugal barrier. $\chi=\sqrt{2}e^2Z_d\sqrt{\mu}Q_p^{-1/2}/\hbar$ is the Coulomb parameter with $\hbar$ and $\mu$ being the reduced Planck constant and the reduced mass of the proton--daughter system, respectively. $Q_p$ is the decay energy of proton. Since $T_{\rm{red}}$ is subtracted from the contribution of the centrifugal barrier, Eq. \eqref{first} contains non the influence of orbital angular momentum. For determining the contribution of the charge of the daughter nucleus in the proton radioactivity half--life, Budaca ${\it et\  al.}$ \cite{Budaca2017} employ the Brown--type empirical formula to study the proton radioactivity half--life based on orbital angular momentum classifications. This formula can be written as
\begin{equation}
\text{log}_{10}T_{1/2}(s)=aZ_{d}^{\beta}Q_{p}^{-1/2}+b,
\end{equation}
where the parameters are as follows
\begin{equation}
\begin{split}
&\beta=0.73,\ \  \text{for}\  l=0,\ 2\  \text{nuclei}\\
&\beta=0.85,\ \  \text{for}\  l=3,\ 5\  \text{nuclei}.
\end{split}
\end{equation}
Here $\beta$ is a particular power parameter of dependence on $Z_{d}$. $l$ is the orbital angular momentum taken away by the emitted proton. To verify the information of proton radioactivity half--life, Sreeja ${\it et\  al.}$ \cite{Sreeja2018} follow the Brown--type empirical formula to propose a new formula with including orbital angular momentum $l$ dependence, it is expressed as
\begin{equation}
\text{log}_{10}T_{1/2}(s)=((a\times l)+b)Z_{d}^{0.8}Q_{p}^{-1/2}+((c\times l)+d),
\end{equation}
where the power dependence on $Z_{d}$ fitting by experiment data is $0.8$. The values of adjustable parameters $a$, $b$, $c$ and $d$ determined by 44 experimental data of proton radioactivity nuclei are as follows
\begin{equation}
\begin{split}
&a=0.0322, \ b=0.8204, \\
&c=-0.1527, \ d=-26.4801.
\end{split}
\end{equation}

Above all, considering the contributions of the charge of daughter nucleus and orbital angular momentum, the G--N law can be applied to the study of proton radioactive half--life. In this work, we propose a two--parameter formula of new G--N law for proton radioactivity. The detailed discussion of the new G--N law is given in Section \ref{section 3}.

\subsection{Spin--parity selection rule and Q value}
The proton emission from nuclei obeys the spin-parity conservation laws
\begin{equation}
J_p=J_d+J_{p^{\iota}}, \ \  \pi_p=\pi_d \pi_{p^{\iota}}(-1)^l,
\end{equation}
where $J_p$, $\pi_p$, $J_d$, $\pi_d$, $J_{p^{\iota}}$ and $\pi_{p^{\iota}}$ are spin and parity values of the parent, daughter and outgoing proton, respectively.
The proton has a nonzero value of spin and positive parity, therefore the minimal value of angular momentum at the proton transition is obtained slightly different from $\alpha$ decay \cite{DENISOV2009815}. It is expressed as
\begin{equation}\label{lmin}
l_{\text{min}}=
\begin{cases}
\Delta_j&\text{for even}\ \Delta_j\  \text{and}\  \pi_p=\pi_d,\\
\Delta_j+1&\text{for even}\ \Delta_j\  \text{and}\  \pi_p\not=\pi_d,\\
\Delta_j&\text{for odd\ }\ \Delta_j\  \text{and}\  \pi_p\not=\pi_d,\\
\Delta_j+1&\text{for odd\ }\ \Delta_j\  \text{and}\  \pi_p=\pi_d,\\
\end{cases}
\end{equation}
where $\Delta_j=\mid J_p-J_d-J_{p^{\iota}}\mid$.

The decay energy $Q_p$ \cite{PhysRevC.96.034619} is generically calculated by
\begin{equation}\label{q-p}
Q_p=\Delta M_p-(\Delta M_d+\Delta M_{p^{\iota}})+k(Z^{\varepsilon}_p-Z^{\varepsilon}_d),
\end{equation}
where $\Delta M_p$, $\Delta M_d$ and $\Delta M_{p^{\iota}}$ are, correspondingly, the mass excess of parent and daughter nuclei and emitted proton. The experimental data of mass excess $\Delta M_p$ and $\Delta M_d$ are taken from the latest evaluated nuclear properties table NUBASE2016 \cite{Audi_2017} and the latest evaluated atomic mass table AME2016 \cite{Huang_2017,Wang_2017}. The term $k(Z^{\varepsilon}_p-Z^{\varepsilon}_d)$ represents the screening effect of the atomic electrons\cite{PhysRevC.72.064613}, where $k = 8.7$ eV, $\varepsilon = 2.517$ for $Z \ge 60$, and $k = 13.6$ eV, $\varepsilon = 2.408$ for $Z < 60$ \cite{HUANG1976243}.

\section{Results and discussions}
\label{section 3}

\begin{figure}[!ht]
\includegraphics[width=8.5cm]{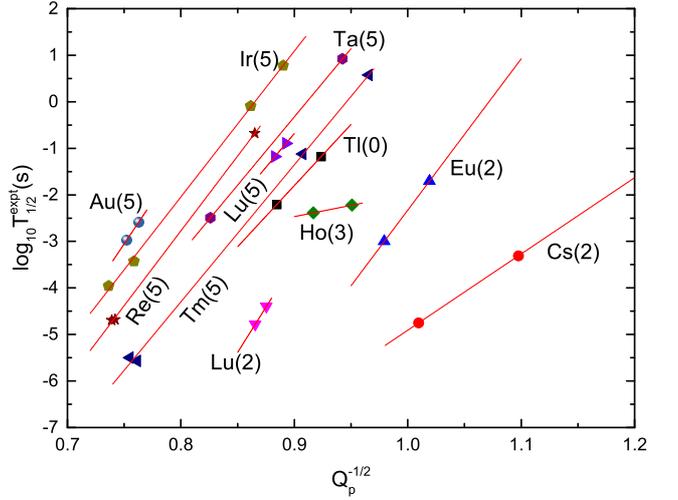}
\caption{(color online) The G--N law plots for different cases of proton radioactivity isotopes using experimental
half-lives and $Q_p$ values. The orbital angular momentum values $l$ taken away by emitted proton are given within the parentheses.}
\label{fig1}
\end{figure}

\begin{figure}[ht]
\includegraphics[width=8.5cm]{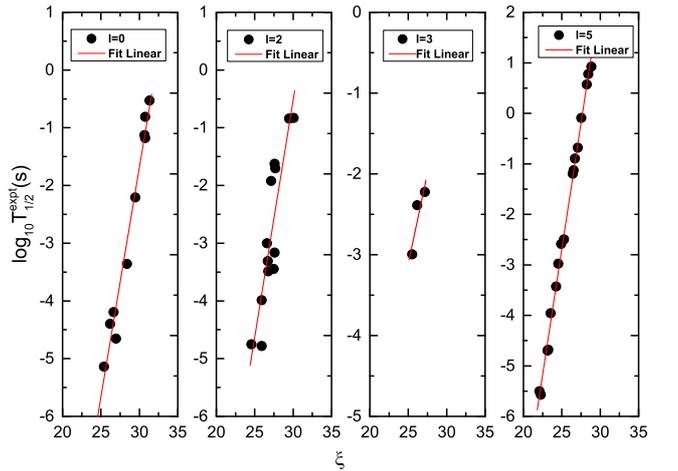}
\caption{(color online) The relationships between the logarithm of the experimental half--life and $\xi$ in the cases of $l=0$, 2, 3 and 5.}
\label{fig2}
\end{figure}

Combining the above empirical linear equations, and considering all the available proton radioactivity experimental data, the G--N law can also be employed in studying the half--life proton radioactivity for the isotopes with being the same orbital angular momentum $l$. For intuitively description about G--N law, we plot the 11 cases (including 27 experimental data) of the straight line between the logarithm of the experimental half--lives of proton radioactivity isotopes and $Q_{p}^{-1/2}$ in Fig. \ref{fig1}. In this figure, all the cases are perfectly following the G--N law and can be expressed as the form of Eq. \eqref{GNLAW}. Committed to extend this linear relationship to all available proton emitters, Budaca \cite{Budaca2017} and Sreeja ${\it et\  al.}$ \cite{Sreeja2018} employ the Brown-type empirical formula \cite{PhysRevC.46.811} to determine the contribution of the charge of the daughter nucleus $Z_d$ i.e. a particular power value of dependence on $Z_{d}$. We plot the relationships between the logarithm of the experimental half--life and $\xi$ in the cases of $l=0$, 2, 3 and 5 in Fig. \ref{fig2}. Where  $\xi =Z_d^{0.8}/\sqrt{Q_p}$ is taken from Ref. \cite{Sreeja2018}. In this circumstance, the Root Mean Square(RMS) deviation of the empirical estimates with respect to the experimental value are found to be 0.301, 0.648, 0.238 and 0.138 corresponding to $l=0$, 2, 3 and 5. The results show that a new G--N law can be obtained by considering the contributions of $Z_d$ and $l$ according to the Geiger--Nuttall--Brown systematics

\begin{figure}[ht]
\includegraphics[width=8.5cm]{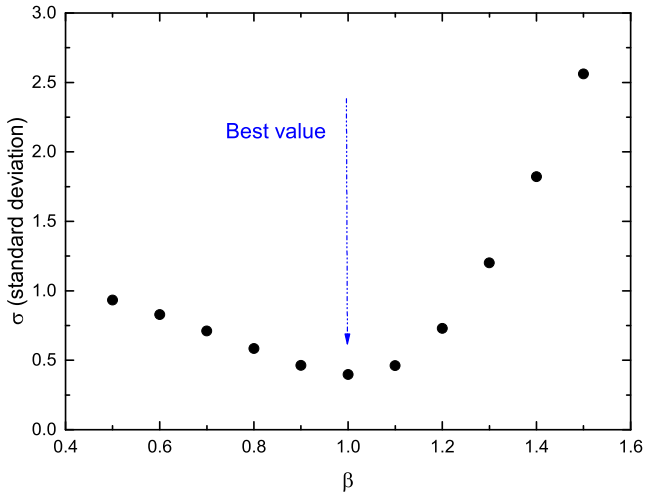}
\caption{(color online) The dependence of the standard deviation on the value of the power $\beta$.}
\label{fig2p}
\end{figure}

Here, being based on the form of G--N law \cite{doi:10.1080/14786441008637156}, considering the contributions of the orbital angular momentum $l$ and the charge of the daughter nucleus $Z_d$ in a unified way and according to the phenomenon in Figs. \ref{fig1} and \ref{fig2}, we propose a simple formula for proton radioactivity half--life. It is expressed as
\begin{equation}
\text{log}_{10}T_{1/2}(s)=a_{\beta}(Z_{d}^{0.8}+l^{\beta})Q_{p}^{-1/2}+b_{\beta}.
\end{equation}
The exponent $\beta$ on the orbital angular momentum $l$ is to determine its effect on the proton radioactivity half--life. Fitting 44 experimental data of the proton radioactivity half-lives in the ground state and isomeric state, we obtain a set of results as shown in Fig. \ref{fig2p}. While the exponent $\beta$ is varied from 0.5 to 1.5. In this figure, the best standard deviation is found to occur at the exponent $\beta = 1.0$. Now, referenced to Brown--type empirical formula, we propose a two--parameters formula for proton radioactivity half--life keeping the exponent of the $Z_d$ and $l$ as 0.8 and 1.0, respectively. It is written as
\begin{equation}\label{newgnl}
\text{log}_{10}T_{1/2}(s)=a(Z_{d}^{0.8}+l)Q_{p}^{-1/2}+b,
\end{equation}
where the parameters corresponding to the best standard deviation are as follow
\begin{equation}
\begin{split}
&a= 0.843,\ \ \ b=-27.194.
\end{split}
\end{equation}
The standard deviation $\sigma$ indicating the deviation between the experimental data and calculated ones can be expressed as
\begin{equation}\label{sigam}
\sigma=\sqrt{\frac{1}{N}\sum\limits_{i=1}^N(\rm{log}_{10}T^{\rm{calc},\it i}_{1/2}-\rm{log}_{10}T^{\rm{expt},\it i}_{1/2})^2}.
\end{equation}
We obtain $\sigma$ between the result calculated by Eq. \eqref{newgnl} and the experimental data is 0.397. It means that the calculations of proton radioactivity half--lives can reproduce the half--lives of experimental data well and differ from the experiment data by a factor of 2.49 on average.

\begin{figure}[!ht]
\includegraphics[width=8.5cm]{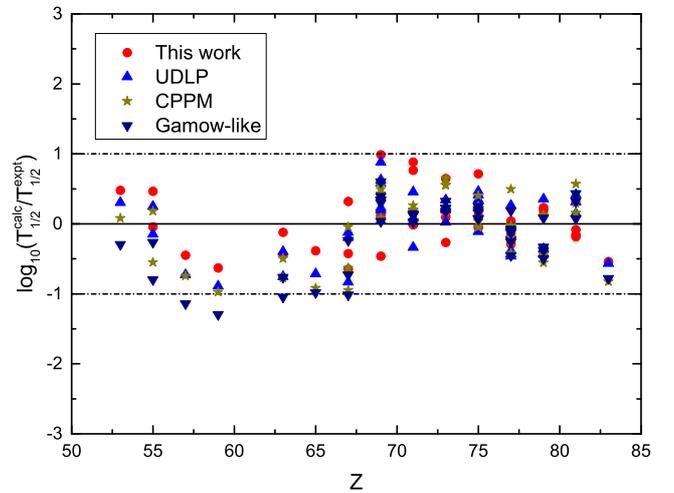}
\caption{(color online) Decimal logarithm deviations between the experimental data of proton radioactivity half-lives and calculations. The circles, upper triangles, pentagrams and lower triangles refer to results obtained by Eq. \eqref{newgnl}, UDLP, the CPPM--Guo2013 and the modified Gamow-like model denoted as Expt, This work, UDLP, CPPM and Gamow--like, respectively.}
\label{fig3}
\end{figure}

\begin{figure}[!ht]
\includegraphics[width=8.5cm]{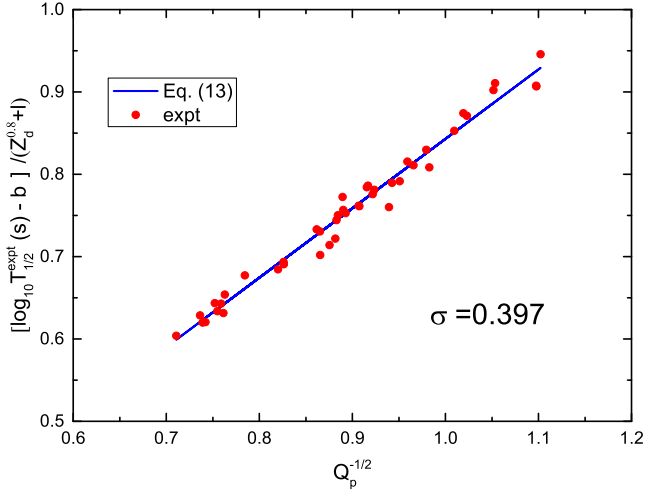}
\caption{(color online) The comparison of the logarithm of the calculated half--lives with the experimental data for proton radioactivity. The line represents the calculated values and the points represent the experimental ones.}
\label{fig4}
\end{figure}

In this work, the experimental proton radioactivity half-lives, spin and parity are taken from Refs. \cite{Zdeb2016,BLANK2008403,Audi_2017}. The decay energies $Q_p$ are given by Eq. \eqref{q-p} with the the mass excess of parent and daughter nuclei taken from Refs. \cite{Huang_2017,Wang_2017} except for $^{117}$La, $^{140,141}$Ho, $^{144}$Tm, $^{150,151}$Lu, $^{159}$Re, $^{159}$Re$^{m}$, $^{164}$Ir, which are taken from Ref. \cite{BLANK2008403}. The values of orbital angular momentum for $^{109}$I and $^{177}$Tl$^m$ are cited from Zedb ${et\ al.}$ \cite{Zdeb2016}. For a comparison, the proton radioactivity half-lives are also evaluated using the universal decay law for proton emission (UDLP) from Qi ${et\ al.}$ \cite{PhysRevC.85.011303}, the CPPM--Guo2013 by Guo ${et\ al.}$ \cite{GUO201354} analyzed from our previous work \cite{Deng2019} and using the modified Gamow--like model from our previous work \cite{Chen_2019}. The detailed results including this work, the UDLP, the CPPM--Guo2013 and the modified Gamow-like model are given in Table \ref{table 1}. In this table, the first three columns present the experimental data of the proton emitter, corresponding $Q_p$ and $l$, respectively. The last five columns are the logarithmic half-lives log$_{10}T_{1/2}$(s) of the experimental data, and obtained using the four theoretical approaches i.e. Eq. \eqref{newgnl}, the UDLP, the CPPM--Guo2013 and the modified Gamow-like model denoted as Expt, This work, UDLP, CPPM and Gamow--like, respectively. In addition, the standard deviations $\sigma$ calculated by Eq. \eqref{sigam} for these calculations i.e. This work, UDLP, CPPM and Gamow--like are listed in Table \ref{table 2} for comparison. The results of the two empirical formulas (UDLP and Eq. \eqref{newgnl}) are better than the results of the two models (CPPM and Gamow--like). Compared to UDLP, our new G--N law slightly improves by $\frac{0.427-0.397}{0.427}=7.0\%$. To obtain further insight into the well of agreement and the systematics of results, we plot the comparison of the calculated half-lives with the experimental data in Fig. \ref{fig3}. It indicates that the ${\rm log}_{10}(T_{1/2}^{\rm calc} /T_{1/2}^{\rm expt})$ values by this work are generally in the range of $\pm 0.5$, corresponding to the proton radioactivity half--lives calculated by Eq. \eqref{newgnl} are within a factor of 2.49. In the Fig. \ref{fig3}, there is a phenomenon that the deviation appears to be uncontinuous around $Z = 68$. It may be caused by an abrupt change in the $Q_p$ values or the nuclear structure of emitters \cite{PhysRevLett.96.072501,FERREIRA2011508}. In order to further demonstrate the significant correlation between the half--lives of proton decay processes and the Q--values of the emitted proton, according to Eq. \eqref{newgnl}, we plot the quantity $[{\rm log}_{10}T_{1/2}^{\rm expt}-b]/(Z_d^{0.8}+l)$ as a function of $Q_p^{-1/2}$ in Fig. \ref{fig4}. It indicates that the linear dependence of the proton radioactivity half--life on $Q_p^{-1/2}$ is obvious within eliminating the contributions of charge number and angular momentum on the proton radioactivity half--life.

\begin{table*}[!htb]
\centering
\caption{A comparison of calculated half-lives of the proton emitters in the ground state and the isomeric state with different theoretical methods and experimental data. The calculations of proton radioactivity half--lives log$_{10}T_{1/2}$(s) displayed as This work, UDLP, CPPM and Gamwo--like are obtained by Eq. \eqref{newgnl}, UDLP, the CPPM--Guo2013 and the Gamow--like model, respectively. The symbol $(m)$ by parent nuclei denote the isomeric state. The experimental $Q_p$ values are given by Eq. \eqref{q-p} and half-lives are taken from Ref. \cite{Audi_2017} except where noted. }
\label{table 1}
\footnotesize
\begin{threeparttable}
\begin{tabular}{lccccccc}
\hline\noalign{\smallskip}
{Nucleus} & $Q_{p}$ (MeV) &  $l$& & & log$_{10}{T_{1/2}}$ (s) \\
\cline{4-8}
  &   &   &  $\text{Expt} $ &  $\text{This work} $ & $\text{UDLP} $ & $\text{CPPM} $ & $\text{Gamow--like} $\\
\noalign{\smallskip}\hline\noalign{\smallskip}
$^{109}\mathrm{I    }$&0.830&   2\tnote{a}& $-$   3.987  &  $-$    3.507    &  $-$    3.684  &  $-$  3.906&  $-$ 4.281   \\
$^{112}\mathrm{Cs   }$&0.830&     2\tnote{b}& $-$   3.310  &  $-$    2.844    &  $-$    3.062  &  $-$   3.128    &  $-$     3.571   \\
$^{113}\mathrm{Cs   }$&0.981&     2\tnote{b}& $-$   4.752  &  $-$    4.796    &  $-$    4.899  &  $-$   5.302    &  $-$     5.547   \\
$^{117}\mathrm{La   }$&0.823\tnote{c}&     2\tnote{c}& $-$   1.623\tnote{c}  &  $-$    2.072    &  $-$    2.350  &  $-$   2.368    &  $-$     2.762   \\
$^{121}\mathrm{Pr   }$&0.901&     2\tnote{b}& $-$   1.921  &  $-$    2.552    &  $-$    2.811  &  $-$   2.894    &  $-$     3.215   \\
$^{130}\mathrm{Eu   }$&1.043&     2\tnote{c}& $-$   3.000  &  $-$    3.121    &  $-$    3.398  &  $-$   3.494    &  $-$     3.761   \\
$^{131}\mathrm{Eu   }$&0.963&     2\tnote{b}& $-$   1.703  &  $-$    2.141    &  $-$    2.458  &  $-$   2.477    &  $-$     2.749   \\
$^{135}\mathrm{Tb   }$&1.193&     3\tnote{b}& $-$   2.996  &  $-$    3.380    &  $-$    3.712  &  $-$   3.912    &  $-$     3.978   \\
$^{140}\mathrm{Ho   }$&1.106\tnote{c}&     3\tnote{c}& $-$   2.222\tnote{c}  &  $-$    1.902    &  $-$    2.342  &  $-$   2.265    &  $-$     2.457   \\
$^{141}\mathrm{Ho^m}$&1.264 &     0\tnote{b}& $-$   5.137  &  $-$    5.783    &  $-$    5.331  &  $-$   5.769    &  $-$     5.865   \\
$^{141}\mathrm{Ho   }$&1.190&     3\tnote{c}& $-$   2.387  &  $-$    2.811    &  $-$    3.220  &  $-$   3.335    &  $-$     3.403   \\
$^{144}\mathrm{Tm   }$&1.725\tnote{c}&     5\tnote{c}& $-$   5.569\tnote{c}  &  $-$    5.216    &  $-$    4.691  &  $-$   5.142    &  $-$     4.965   \\
$^{145}\mathrm{Tm }$&1.754  &     5\tnote{b}& $-$   5.499  &  $-$    5.401    &  $-$    4.871  &  $-$   5.415    &  $-$     5.164   \\
$^{146}\mathrm{Tm   }$&0.904&   0\tnote{b}& $-$   0.810  &  $-$    1.272    &  $-$    0.610  &  $-$   0.315    &  $-$     0.773   \\
$^{146}\mathrm{Tm^m}$&1.214 &     5\tnote{c}& $-$   1.125  &  $-$    0.999    &  $-$    0.896  &  $-$   0.695    &  $-$     0.776   \\
$^{147}\mathrm{Tm   }$&1.133&     2\tnote{b}& $-$   3.444  &  $-$    2.455    &  $-$    2.859  &  $-$   2.911    &  $-$     3.051   \\
$^{147}\mathrm{Tm^m}$&1.072 &     5\tnote{b}& \ \ \ 0.573   & \ \ \ 0.681      & \ \ \ 0.614    & \ \ \ 1.001      & \ \ \  0.874   \\
$^{150}\mathrm{Lu^m }$&1.305&     2\tnote{c}& $-$   4.398  &  $-$    3.633    &  $-$    4.734  &  $-$   4.206    &  $-$     4.367   \\
$^{150}\mathrm{Lu }$&1.283\tnote{c}  &     5\tnote{c}& $-$   1.194\tnote{c}  &  $-$    1.199    &  $-$    1.113  &  $-$   0.931    &  $-$     1.044   \\
$^{151}\mathrm{Lu^m }$&1.335&     2\tnote{b}& $-$   4.783  &  $-$    3.899    &  $-$    4.327  &  $-$   4.582    &  $-$     4.662   \\
$^{151}\mathrm{Lu   }$&1.255\tnote{c}&     5\tnote{c}& $-$   0.896\tnote{c}  &  $-$    0.911    &  $-$    0.863  &  $-$   0.699    &  $-$     0.767   \\
$^{155}\mathrm{Ta }$&1.466  &     5\tnote{b}& $-$   2.495  &  $-$    2.397    &  $-$    2.269  &  $-$   2.321    &  $-$     2.267   \\
$^{156}\mathrm{Ta   }$&1.036&     2\tnote{b}& $-$   0.828  &  $-$    0.180    &  $-$    0.624  &  $-$   0.279    &  $-$     0.649   \\
$^{156}\mathrm{Ta^m}$&1.126 &     5\tnote{b}&\ \ \  0.924   &\ \ \  1.101      &\ \ \  0.947    &\ \ \  1.479      &\ \ \   1.248   \\
$^{157}\mathrm{Ta   }$&0.956&     0\tnote{b}& $-$   0.529  &  $-$    0.797    &  $-$    0.188  &        0.122     &  $-$    0.305   \\
$^{159}\mathrm{Re   }$&1.816\tnote{c}&     5\tnote{c}& $-$   4.678\tnote{c}  &  $-$    4.493    &  $-$    4.268  &  $-$   4.639    &  $-$     4.428   \\
$^{159}\mathrm{Re^m}$&1.831\tnote{c} &     5\tnote{c}& $-$   4.695\tnote{c}  &  $-$    4.586    &  $-$    4.355  &  $-$   4.741    &  $-$     4.524   \\
$^{160}\mathrm{Re   }$&1.286&     2\tnote{c}& $-$   3.164  &  $-$    2.450    &  $-$    2.939  &  $-$   2.915    &  $-$     3.090   \\
$^{161}\mathrm{Re }$&1.216&       0\tnote{b}& $-$   3.357  &  $-$    3.277    &  $-$    2.895  &  $-$   2.953    &  $-$     3.152   \\
$^{161}\mathrm{Re^m  }$&1.336&    5\tnote{b}& $-$   0.680  &  $-$    0.729    &  $-$    0.789  &  $-$   0.579    &  $-$     0.601   \\
$^{164}\mathrm{Ir }$&1.844\tnote{c}&       5\tnote{c}& $-$   3.959\tnote{c}  &  $-$    4.247    &  $-$    4.114  &  $-$   4.376    &  $-$     4.213   \\
$^{165}\mathrm{Ir^m  }$&1.737&    5\tnote{b}& $-$   3.430  &  $-$    3.550    &  $-$    3.472  &  $-$   3.685    &  $-$     3.502   \\
$^{166}\mathrm{Ir }$&1.177&       2\tnote{b}& $-$   0.842  &  $-$    0.801    &  $-$    1.303  &  $-$   1.036    &  $-$     1.294   \\
$^{166}\mathrm{Ir^m  }$&1.347&    5\tnote{b}& $-$   0.091  &  $-$    0.344    &  $-$    0.475  &  $-$   0.136    &  $-$     0.215   \\
$^{167}\mathrm{Ir }$&1.087&       0\tnote{b}& $-$   1.128  &  $-$    1.347    &  $-$    0.865  &  $-$   0.631    &  $-$     0.940   \\
$^{167}\mathrm{Ir^m  }$&1.262&    5\tnote{b}& \ \ \ 0.778   &\ \ \  0.546      &\ \ \  0.348    &\ \ \  0.758      &\ \ \   0.683   \\
$^{170}\mathrm{Au  }$&1.487&      2\tnote{b}& $-$   3.487  &  $-$    3.254    &  $-$    3.845  &  $-$   3.927    &  $-$     3.984   \\
$^{170}\mathrm{Au^m  }$&1.767&    5\tnote{b}& $-$   2.975  &  $-$    3.330    &  $-$    3.333  &  $-$   3.442    &  $-$     3.307   \\
$^{171}\mathrm{Au   }$&1.464&     0\tnote{b}& $-$   4.652  &  $-$    4.460    &  $-$    4.298  &  $-$   4.527    &  $-$     4.569   \\
$^{171}\mathrm{Au^m}$&1.718&      5\tnote{b}& $-$   2.587  &  $-$    2.992    &  $-$    3.026  &  $-$   3.144    &  $-$     2.966   \\
$^{176}\mathrm{Tl  }$&1.278&      0\tnote{b}& $-$   2.208  &  $-$    2.361    &  $-$    2.059  &  $-$   1.909    &  $-$     2.133   \\
$^{177}\mathrm{Tl   }$&1.172&     0\tnote{b}& $-$   1.178  &  $-$    1.263    &  $-$    0.863  &  $-$   0.610    &  $-$     0.855   \\
$^{177}\mathrm{Tl^m  }$&1.979&   6\tnote{a}& $-$   3.459  &  $-$    3.643    &  $-$    3.045  &  $-$   3.305    &  $-$     3.025   \\
$^{185}\mathrm{Bi^m  }$&1.625&    0\tnote{b}& $-$   4.192  &  $-$    4.730    &  $-$    4.759  &  $-$   5.017    &  $-$     4.971   \\

\noalign{\smallskip}\hline
\end{tabular}
\begin{tablenotes}
\item[\makelabel{a}] Taken from Ref. \cite{Zdeb2016}.
\end{tablenotes}
\begin{tablenotes}
\item[\makelabel{b}] Taken from Ref. \cite{Audi_2017}.
\end{tablenotes}
\begin{tablenotes}
\item[\makelabel{c}] Taken from Ref. \cite{BLANK2008403}.
\end{tablenotes}
\end{threeparttable}
\end{table*}

\begin{table}[]
\caption{The standard deviations $\sigma$ of the calculations from Eq. \eqref{newgnl}, UDLP, CPPM and Gamow--like with respect to the experimental data.}
\label{table 2}
\begin{tabular}{ccccc}
\hline\noalign{\smallskip}
Model&This work&UDLP&CPPM&Gamow--like\\
\noalign{\smallskip}\hline\noalign{\smallskip}
$\sigma$&0.397&0.427&0.472&0.501\\
\noalign{\smallskip}\hline
\end{tabular}
\end{table}

\begin{figure}[!ht]
\includegraphics[width=8.5cm]{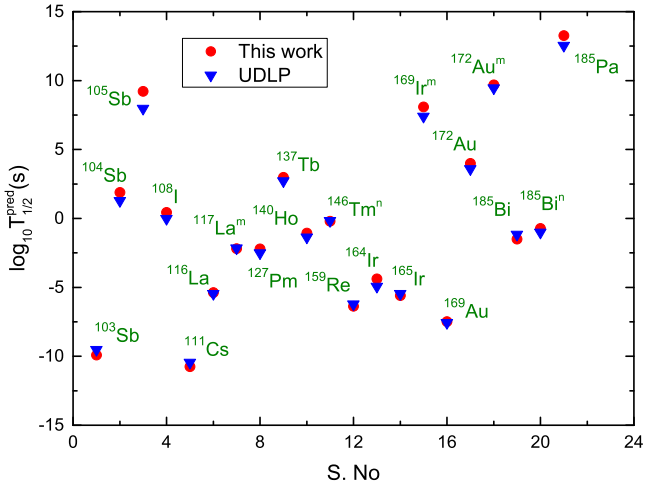}
\caption{(color online) Comparison of the predicted proton radioactivity half-lives using the new G--N law and UDLP. The circles and triangles refer to results obtained by Eq. \eqref{newgnl} and UDLP.}
\label{fig5}
\end{figure}

\begin{table*}[!htp]
\centering
\caption{A comparison of predicted proton radioactivity half-lives of nuclei in region $51\leq Z \leq91$ using Eq. \eqref{newgnl} and UDLP. The symbol $(m, n)$ by parent nuclei denote the isomeric state. $'()'$ means uncertain spin and / or parity. $'\#'$ means values estimated from trends in neighboring nuclides with the same $Z$ and $N$ parities.}
\label{table 3}

\begin{tabular}{clcccccc}
\hline\noalign{\smallskip}
{S.No} &{Nucleus} & $Q_{p}$ (MeV) &${J^{\pi}_{p}}\to{J^{\pi}_{d}}$ & $l_{\text{min}}$& & log$_{10}{T_{1/2}}$ (s) \\
\cline{6-8}
  &  &   &   &   &  $\text{This work} $ & $\text{UDLP} $ & $\text{Expt} $  \\
\noalign{\smallskip}\hline\noalign{\smallskip}
1&$^{103}\mathrm{Sb     }$&1.469   &${ 5/2^+\#  	}\to{ 0^+     }$&   2& $-$    9.902    &  $-$   9.515      &                    \\
2&$^{104}\mathrm{Sb     }$&0.519   &${ ()       	}\to{ 5/2^+\#  }$&   2&\ \ \   1.890    &\ \ \   1.278      &\ \ \ $>$  0.827   \\
3&$^{105}\mathrm{Sb     }$&0.331   &${ (5/2+) 	}\to{ 0+     }$&   2&\ \ \   9.216    &\ \ \   7.980      & \ \ \ $>$  3.049       \\
4&$^{108}\mathrm{I      }$&0.610   &${ 1^+\#    	}\to{ 5/2^+\#  }$&   2&\ \ \   0.433    &  $-$   0.024     &\ \ \ $>$ 0.556   \\
5&$^{111}\mathrm{Cs     }$&1.820   &${ 3/2^+\#  	}\to{ 0^+     }$&   2& $-$    10.751   &  $-$   10.445     &                    \\
6&$^{116}\mathrm{La     }$&1.091   &${ ()       	}\to{ 5/2^+\#  }$&   2& $-$    5.373    &  $-$   5.456      &                   \\
7&$^{117}\mathrm{La^m   }$&0.951   &${ (9/2^+) 	}\to{ 0^+     }$&   4& $-$    2.191    &  $-$   2.155      &  $\approx-$ 1.989         \\
8&$^{127}\mathrm{Pm     }$&0.922   &${ 5/2^+\#  	}\to{ 0^+     }$&   2& $-$    2.209    &  $-$   2.514      &                   \\ 
9&$^{129}\mathrm{Pm     }$&0.152   &${ (5/2^-) 	}\to{ 0^+     }$&   3&\ \ \   36.533   &\ \ \   33.398     &                   \\ 
10&$^{137}\mathrm{Tb     }$&0.843   &${ 11/2^-\# 	}\to{ 0^+     }$&   5&\ \ \   2.977    & \ \ \  2.717      &                   \\ 
11&$^{140}\mathrm{Ho     }$&1.103   &${ 8^+\#    	}\to{ (7/2^+) }$&   4& $-$    1.069    &  $-$   1.364      &                   \\ 
12&$^{146}\mathrm{Tm^n   }$&1.144   &${ (10^+)  	}\to{ 11/2^-\# }$&   5& $-$    0.204    &  $-$   0.176      &                   \\ 
13&$^{159}\mathrm{Re     }$&1.605   &${ 1/2^+\#  	}\to{ 0^+     }$&   0& $-$    6.377    &  $-$   6.223      &                   \\ 
14&$^{164}\mathrm{Ir     }$&1.576   &${ 2^-\#    	}\to{ 7/2^-   }$&   2& $-$    4.387    &  $-$   4.947      &                   \\ 
15&$^{165}\mathrm{Ir     }$&1.556   &${ 1/2^+\#  	}\to{ 0^+     }$&   0& $-$    5.593    &  $-$   5.455      &                   \\ 
16&$^{169}\mathrm{Ir^m   }$&0.780   &${ (11/2^-)	}\to{ 0^+     }$&   5& \ \ \  8.088    &\ \ \   7.404      &                   \\ 
17&$^{171}\mathrm{Ir ^m  }$&0.402   &${ (11/2^-)	}\to{ 0^+     }$&   5& \ \ \ 21.952    & \ \ \  20.396     &                  \\  
18&$^{169}\mathrm{Au     }$&1.947   &${ 1/2^+\#  	}\to{ 0^+     }$&   0& $-$    7.476    &  $-$   7.569      &                    \\
19&$^{172}\mathrm{Au     }$&0.877   &${ (2^-)   	}\to{ 7/2^-   }$&   2&\ \ \    3.991   &\ \ \   3.586      & \ \ \  $>$0.146    \\
20&$^{172}\mathrm{Au  ^m }$&0.627   &${ (9^+)   	}\to{ 13/2^+  }$&   2&\ \ \    9.692   &\ \ \   9.448      & $>-$ 0.260        \\
21&$^{185}\mathrm{Bi     }$&1.540   &${ 9/2^-\#  	}\to{ 0^+     }$&   5& $-$    0.721    &  $-$   1.019      &                   \\ 
22&$^{185}\mathrm{Bi^n   }$&1.720   &${ 13/2^+\# 	}\to{ 0^+     }$&   6& $-$    1.502    &  $-$   1.161      &                   \\ 
23&$^{211}\mathrm{Pa     }$&0.751   &${ 9/2^-\#  	}\to{ 0^+     }$&   5&\ \ \    13.268  &\ \ \   12.545     &                   \\

\noalign{\smallskip}\hline
\end{tabular}
\end{table*}

In the following, as a promotion, we extend the new G--N law to predict the proton radioactivity half-lives of 23 nuclei in region $51 \leq Z \leq 91$, whose proton radioactivity is energetically allowed or observed but not yet quantified in NUBASE2016 \cite{Audi_2017}. The results are listed in table \ref{table 3}. In this table, the first five columns present the serial number, proton emitter, corresponding decay energy $Q_p$, the transferred minimum orbital angular momentum $l_{\rm min}$ and the spin--parity transformed from the parent to daughter nuclei, respectively. The last three columns are the logarithmic half-lives log$_{10}T_{1/2}$(s) of Eq. \eqref{newgnl}, the UDLP and the experimental data, denoted as This work, UDLP and Expt, respectively. The spin and parity are taken from the NUBASE2016 \cite{Audi_2017}, the experimental data of mass excess $\Delta M_p$ and $\Delta M_d$ are taken from the AME2016 \cite{Huang_2017,Wang_2017}. Proton radioactive half-life is very sensitive to orbital angular momentum and decay energy, however most of the information ${J^{\pi}_{p}}\to{J^{\pi}_{d}}$ in Table \ref{table 3} is not unique. In order to reduce the error in predicting half--life, the minimum orbital angular momentum $l_{\rm min}$ and the decay energy $Q_p$ are corresponding calculated by Eqs. \eqref{lmin} and \eqref{q-p}. The $l_{\rm min}$ for uncertain nuclei are the theoretically established minimum orbital angular momentum. Some of half-life information log$_{10}{T_{1/2}}^{\text{Expt}}(s)$ provided by NUBASE2016 is as $^{104}{\rm Sb} > 0.827$,  $^{105}{\rm Sb} > 3.049$, $^{108}{\rm I} > 0.556$, $^{117}{\rm La}^m \approx -1.989$, $^{172}{\rm Au} > 0.146$ and$^{172}{\rm Au}^m > -0.260$. The above results predicted by our method are all within the range except $^{108}{\rm I}$. The reason for the deviation of $^{108}{\rm I}$ may be that its orbital angular momentum is not accurate. As for $^{129}{\rm Pm}$ and $^{171}{\rm Ir}^m$, the reason for the high order magnitude of the calculated results may be that $Q_p$ is too small. For a comparison, the proton radioactivity half-lives of these nuclei are also predicted using UDLP. The comparison results are presented in Fig. \ref{fig5}. In this figure, the proton emitters are given in the same order as Table \ref{table 3} except for $^{129}{\rm Pm}$ and $^{171}{\rm Ir}^m$ being not listed. It indicates that the predictions of our new G--N law are consistent with UDLP.

\section{Summary}
\label{section 4}
To summarize, according to the form of Geiger--Nuttall law, considering the contributions of the orbital angular momentum $l$ and the charge of the daughter nuclei $Z_d$ in a unified way, we propose the new G--N law i.e. a two--parameter empirical formula of half--life for proton radioactivity. By fitting 44 experimental data of proton emitters in the ground state and isomeric state, we obtain a set of parameters. The results show well agreement between the experimental data and the calculated values. In this sense, we extend the new G--N law to predict proton radioactivity half-lives for $51 \leq Z \leq 91$ nuclei, whose proton radioactivity is energetically allowed or observed but not yet quantified in NUBASE2016. Furthermore, the predictions of our new G--N law are consistent with UDLP. The formula is very convenient to study proton radioactivity, and could be useful for future experiments.

\noindent
This work is supported in part by the National Natural Science Foundation of China (Grants No. 11205083 and No. 11505100), the Construct Program of the Key Discipline in Hunan Province, the Research Foundation of Education Bureau of Hunan Province, China (Grant No. 15A159 and No. 18A237), the Natural Science Foundation of Hunan Province, China (Grants No. 2015JJ3103 and No. 2015JJ2121), the Innovation Group of Nuclear and Particle Physics in USC, the Shandong Province Natural Science Foundation, China (Grant No. ZR2015AQ007), the Hunan Provincial Innovation Foundation For Postgraduate (Grant No. CX20190714), and the National Innovation Training Foundation of China (Grant No. 201910555161).

%
%

\end{document}